\documentclass[a4paper,11pt,fleqn]{article}
\usepackage[latin1]{inputenc}
\usepackage[T1]{fontenc}
\usepackage{amsfonts}
\usepackage{amssymb}
\usepackage{latexsym}
\usepackage{longtable}
\usepackage{array}
\usepackage{amsmath}
\usepackage{graphicx}
\usepackage{float}

\numberwithin{equation}{section}

\makeatletter
\renewcommand{\section}{\@startsection{section}{1}{0mm}   {\baselineskip}{0.5\baselineskip}{\underline\scshape\normalfont\large\textbf}}
\makeatother
\makeatletter
\renewcommand{\subsection}{\@startsection{subsection}{2}{0mm}   {\baselineskip}{1.5\baselineskip}{\normalfont\large\scshape\textbf}}
\makeatother

\renewcommand{\Re}{\textrm{Re}}
\newcommand{\Trev}{T_{\rm{rev}}}
\newcommand{\Tcl}{T_{\rm{cl}}}

\title{\textsc{Reconstruction and location of fractional revivals of coherent state wave-packets for potentials associated with exceptional $X_m$ Jacobi-polynomials}}
\author{\textbf{Sid-Ahmed YAHIAOUI\thanks{email address: \texttt{s$\_$yahiaoui@univ-blida.dz} and \texttt{sid.phy@gmail.com}}}\quad and\quad\textbf{Mustapha BENTAIBA}\thanks{email address: \texttt{bentaiba@univ-blida.dz}}\\[1mm]
\textit{LPTHIRM, d\'epartement de physique, facult\'e des sciences,}\\[-3mm] \textit{universit\'e Sa\^ad DAHLAB de Blida 1,}\\[-3mm]\textit{B.P. 270 Route de Soum\^aa, 09000 Blida, Algeria}\\
}

\begin{document}

\maketitle

\begin{abstract}
\noindent Gazeau-Klauder coherent states of the extended trigonometric Scarf potential, underlying the quadratic energy spectrum and associated with Jacobi-type $X_m$ exceptional orthogonal polynomials $\mathcal P_n^{(a,b,m)}(x)$, are constructed. The temporal evolution of wave-packet coherent states are performed by means of an autocorrelation function and the full revival properties are investigated in the usual time-domain analysis. This latter seems to be less useful for describing the fractional revivals due to the complicated nature of coherent wave-packet. Fortunately the autocorrelation function revels a little signature of fractional revivals at the vicinity of quarters of the revival time $\Trev$ due to the quadratic energy spectrum and the use of the wavelet-based time-frequency analysis of the autocorrelation function provides an analytical and numerical observation of the fractional revivals at different orders of the system.\\[1.5mm]
\end{abstract}
\noindent Submitted to: \textit{J. Phys. A: Math. Theor.}
\vspace{1mm}

\section{Introduction}%

\noindent Coherent states (CS) were first introduced by Schr\"odinger \cite{1} in the context of quantum mechanics in order to provide a close connection between classical and quantum formalism. They are defined as a specific superposition of the eigenstates of the harmonic oscillator \cite{2,3,4} and, conventionally, there are three ways to define them:
\begin{description}
  \item[(HOCS1-)] they are a set of eigenstates of the annihilation operator $\hat a$ (Glauber's approach),
  \item[(HOCS2-)] they are regarded as those states obtained by the application of the displacement operator $\hat D(z)=\exp\{z\hat a^\dagger-z^\ast \hat a\}$ upon the vacuum state (Klauder's approach), and
  \item[(HOCS3-)] they minimize the Heisenberg position-momentum uncertainty relation (Schr\"odinger's approach).
\end{description}

\indent Later several proposals for generalizing CS in the context of Lie group structures are suggested \cite{5,6,7,8}, among them we find those constructed by Nieto and Simmons \cite{9}, and also by Gazeau and Klauder \cite{10}. The former should be considered only in the semiclassical limit which saturate a generalized uncertainty relation, while the latter are initially constructed without resort to a Lie algebra symmetry and parameterized by two real parameters; i.e. $0\leq J\leq+\infty$ and $-\infty\leq\gamma\leq+\infty$, for models with one degree of freedom with discrete and/or continuous spectra. They are defined by:
\begin{equation}\label{1.1}
  |J,\gamma\rangle=\frac{1}{\mathcal{N}(J)}\sum_{n=0}^{\infty} \frac{J^{n/2}\exp\{-i\gamma e_{n}\}}{\sqrt{\rho_n}}|n\rangle,
\end{equation}
and $\rho_n$ denotes the moments of a probability distribution $\rho(x)$
\begin{eqnarray}\label{1.2}
  \rho_n &=& \int_0^R x^n\rho(x)dx=\prod_{i=1}^n e_i,\quad\quad\rho_0=1,
\end{eqnarray}
where $0<J\leq R=\lim\sup_{n\rightarrow+\infty}\sqrt[n]{\rho_n}$ and $R$ refers to the radius of convergence. Here $e_i$ are the spectrum of the Hamiltonian under consideration arranged in a suitable form.\\
\indent In Gazeau-Klauder formalism, a coherent state should satisfy the following criteria: (i) continuity of labeling, (ii) temporal stability, (iii) resolution of identity, and (iv) action identity. As a consequence, several new classes of CS were investigated, such as those often called "nonclassical" which have attracted a lot of attention. Nonclassical effects are characterized by some purely quantum mechanical properties \cite{11}, such as squeezing, photon antibunching, sub-Poissonian photon statistics, and so on.\\
\indent However in certain quantum systems with nonlinear (particulary, the quadratic) dependence energy spectra, the temporal evolution of wave-packets lead to quantum and fractional revivals \cite{12,13,14}. Such revivals have been studied extensively in Rydberg atoms \cite{15,16,17,18,19}, optical parametric oscillators \cite{20,21}, the Jaynes-Cummings model \cite{22,23}, as well as for systems endowed within position-dependent mass \cite{24,25}. Recently, revival dynamics have been put forward, particulary, for studying the low dimensional systems \cite{26,27}, the quantum phase transitions \cite{28}, the measure of nonclassicality by means of Fisher information \cite{29}, and also have been related to time-dependent $q$-deformed CS under generalized uncertainty relations \cite{30}. The full revival arises when the wave-packet spreads inside the potential and reconstructs itself during its evolution after a certain time. This phenomenon is observed at long-time, called the revival time $\Trev$ ($\Trev\gg\Tcl$), on which the wave-packet is relocalized in the form of a quantum revivals. The fractional revival can be considered as an additional temporal structures and found at times equal to rational fractions of the revival time $r\Trev/l$ with smaller periodicity $\Tcl/l$ [$r/l$ being an irreducible fraction]. This phenomenon is then subjected to partial reconstruction of the initial wave-packet, which are distributed among "$l$ mini-packets" of nearly constant value with $1/l$ of the total probability density.\\
\indent In order to reproduce these revival structures, the autocorrelation function \cite{31,32} is considered as widely used technique for describing the revival dynamics of wave-packets and is usually defined by the overlap integral
\begin{eqnarray}\label{1.3}
  A(t) &=& \langle\psi(x,0)|\psi(x,t)\rangle\nonumber \\
  &=& \int_{-\infty}^{+\infty}\psi^\ast(x,0)\psi(x,t)dx,
\end{eqnarray}
where squared modulus of $A(t)$, $|A(t)|^2$, describes the probability of finding the system at time $t$ in the vicinity of the initial wave-packet $\psi(x,0)$. The time-domain analysis is a mathematical tool for the autocorrelation function, which use Fourier transform, to reveal perfect quantum revivals for systems whose energy spectrum is purely quadratic in the quantum number \cite{33}, but usually will fail to determine precisely the order of the fractional revivals since they appear at specific instants of time.\\
\indent To overcome this problem and put into evidence these fractional revivals, an efficient method is to apply the continuous wavelet transform (CWT) \cite{34,35,36,37,38} which is powerful for multi-resolution local spectrum of non-stationary signals. The CWT is a two-dimensional function of the scale $s$ and the time shift $\tau$ referred to as the timescale wavelet representation and equivalent to a powerful time-frequency analysis; the scale parameter $s$ stretches ($s>1$) or compresses ($0<s<1$) the wavelet till the required resolution while the translation parameter $\tau$ shifts the wavelet to the desired location. The CWT of a function $f(t)\in\mathbb{L}^2$ (the vector space of measurable) is defined as a decomposition of $f(t)$ into a set of basis functions $h_{s,\tau}(t)$, often called the wavelets, and generated from a simple basic wavelet, known as the mother wavelet, given by
\begin{equation}\label{1.4}
    h_{s,\tau}(t)=\frac{1}{\sqrt s}\,h\bigg(\frac{t-\tau}{s}\bigg),
\end{equation}
and, using \eqref{1.4}, CWT may be regarded as an inner product between the signal, $f(t)$, and a set of kernel functions (wavelets)
\begin{eqnarray}\label{1.5}
  W_f(s,\tau)&=&\int f(t)h^\ast_{s,\tau}(t)dt\nonumber\\
             &=&\frac{1}{\sqrt s}\int f(t)h^\ast\bigg(\frac{t-\tau}{s}\bigg)dt.
\end{eqnarray}
\indent However the choice of the kernel function is not specific which represents the main difference between the CWT and other well-known transforms such as Fourier, Laplace and Mellin \cite{31}, and so on. Therefore in order to perform calculations, it is important to know what wavelet is used in \eqref{1.4}.\\
\indent In the present paper, we shall construct coherent states \emph{à la} Gazeau-Klauder for the Hermitian extended Scarf I potential associated with Jacobi-type $X_m$ exceptional orthogonal polynomials (EOP). These latter are known as the solution of the second-order Strum-Liouville eigenvalues problem with rational coefficients \cite{39}. It was seen that although the evolution of the system is effectively governed by one time scale, here $\Trev$, its structure at times of fractional revival is not necessarily simple to be construct. Thus our main aim of this work is the reconstruction and location of fractional revivals corresponding to the Hermitian extended Scarf I potential, by choosing the Morlet wavelet \cite{34,35,36} as a specific kernel function and evaluating the energy density function $\mathcal E(s,\tau)=|W_{|A|^2}(s,\tau)|^2$ for the squared modulus of the autocorrelation function, using a wavelet-based time-frequency analysis rather than the time-domain analysis of the autocorrelation function. We show that the occurrence and the location of fractional revivals for our system is clearly better indicated by CWT than by the standard autocorrelation function.\\
\indent The organization of this paper is as follows: in Section 2 we describe some generic features of the Hermitian extended Scarf I potential in context of $X_m$ Jacobi EOP, and construct their associated Gazeau-Klauder coherent states. The quantum revival dynamics of the deduced coherent states are studied in Section 3 through the construction of their autocorrelation function. In Section 4, we use the CWT of the autocorrelation function to reconstruct and localize fractional revivals for the coherent wave-packet. Finally, Section 5 is devoted to our conclusion.

\section{Gazeau-Klauder coherent states for the Hermitian extended Scarf I potential associated with $X_m$ Jacobi EOP}%

\noindent In recent work Midya and Roy \cite{39} studied an infinite families of the exactly solvable Hermitian as well as non-Hermitian Hamiltonians whose bound-state wave-functions are given in terms of exceptional $X_m$ Jacobi orthogonal polynomials
\begin{eqnarray}\label{2.1}
  \mathcal{P}_n^{(a,b,m)}(x)&=&(-1)^m\bigg[\frac{a+b+j+1}{2(a+j+1)}(x-1)P_m^{(-a-1,b-1)}(x)P_{j-1}^{(a+2,b)}(x)\nonumber \\
                            && +\frac{a-m+1}{a+j+1}P_m^{(-a-2,b)}(x)P_{j}^{(a+1,b-1)}(x)\bigg],\qquad\qquad\qquad(j=n-m\geq0)
\end{eqnarray}
with $-1\leq x\leq+1$, and $\mathcal{P}_n^{(a,b,0)}(x)=P_n^{(a,b)}(x)$ and $P$ are the standard Jacobi polynomials \cite{40}. The real parameters $a$ and $b$ verify the following conditions simultaneously
\begin{eqnarray*}
  (i)\quad b &\neq&0,\,a,\,a-b-m+1\not\in\{0,1,\ldots,m-1\}\nonumber\\
  (ii)\quad a&>&m-2,\quad\textrm{sgn}(a-m+1)=\textrm{sgn}(b),
\end{eqnarray*}
where $\textrm{sgn}(\alpha)=\alpha/|\alpha|$, or equivalently
\begin{eqnarray}\label{2.2}
  (i)\quad &-1<b<0& \textrm{and}\quad m-2<a<m-1\nonumber\\
  (ii)\quad &b>0&   \textrm{and}\quad a>m-1.
\end{eqnarray}
\indent As a direct application two cases are considered in \cite{39}, where one of them concerns the infinite numbers of rationally extended real and trigonometric Scarf potential given by:
\begin{eqnarray}\label{2.3}
  V^{(m)}(x)&=&\frac{k^2(2a^2+2b^2-1)}{4}\sec^2kx-\frac{k^2(b^2-a^2)}{2}\sec kx\tan kx-2k^2m(a-b-m+1)\nonumber \\
            &&-k^2(a-b-m+1)(a+b+(a-b+1)\sin kx)\frac{P_{m-1}^{(-a,b)}(\sin kx)}{P_{m}^{(-a-1,b-1)}(\sin kx)}\nonumber\\
            &&+\frac{k^2(a-b-m+1)^2\cos^2kx}{2}\Bigg[\frac{P_{m-1}^{(-a,b)}(\sin kx)}{P_{m}^{(-a-1,b-1)}(\sin kx)}\Bigg]^2,
\end{eqnarray}
where $-\pi/(2k)<x<\pi/(2k)$ and $k\neq0$. Both eigenvalues and eigenfunctions can be obtained analytically (with $\hbar=2M=1$)
\begin{eqnarray}
  E_n^{(m)} &=& \frac{k^2}{4}(2n-2m+a+b+1)^2,\label{2.4} \\
  \psi_n^{(m)}(x) &=& N_n^{(m)}\frac{(1-\sin kx)^{a/2+1/4}(1+\sin kx)^{b/2+1/4}}{P_{m}^{(-a-1,b-1)}(\sin kx)}\,\mathcal{P}_n^{(a,b,m)}(\sin kx),\label{2.5}
\end{eqnarray}
where $n=m,\,m+1,\,m+2,\ldots$ and the normalization constant is given by
\begin{eqnarray}\label{2.6}
  N_n^{(m)}=\sqrt{k\frac{(2n-2m+a+b+1)(n-m+a+1)^2\Gamma(n-m+1)\Gamma(n-m+a+b+1)}{2^{a+b+1}(n-2m+a+1)\Gamma(n-m+a+2)\Gamma(n+b+1)}}.
\end{eqnarray}
\indent It is to be noted that for $m=0$, the potential \eqref{2.3} is reduced to the well-known Scarf I potential. The potential given in \eqref{2.3} are infinite in number due to the fact that each integer value of $m\geq0$ gives rise to a new exactly solvable potential and each of them become singular at the zeros of the standard Jacobi polynomials, $P_{m}^{(-a-1,b-1)}(\sin kx)$, inside the interval $-\pi/(2k)<x<\pi/(2k)$. These singularities can be avoided by applying the restriction \eqref{2.2}.\\
\indent Let us now adapt the material developed above to construct the Gazeau-Klauder coherent states for the extended real Scarf I potential \eqref{2.3}. For the case at hand we restrict ourself to the first condition of \eqref{2.2}. To this end, let us consider the dynamics of the wave-packet in the extended real Scarf I potential with eigenvalues \eqref{2.4}:
\begin{eqnarray}\label{2.7}
  E_n^{(m)} &=& \frac{k^2}{4}(2n-2m+a+b+1)^2\nonumber \\
            &=& \omega\,e_n^{(m)},\qquad\qquad\textrm{with}\qquad\qquad e_n^{(m)}=(2n-2m+a+b+1)^2
\end{eqnarray}
with the restriction $\omega=k^2/4$. The Gazeau-Klauder coherent states \eqref{1.1} for this system is given by
\begin{eqnarray}\label{2.8}
  \Xi^{(m)}(x;J,\gamma)=\frac{1}{\mathcal{N}(J)}\sum_{n=0}^{+\infty}\frac{J^{n/2}\exp\{-i\gamma e_n^{(m)}\}}{\sqrt{\rho_n}}\,\psi_n^{(m)}(x),
\end{eqnarray}
where $\gamma=\omega t$. For the coherent states \eqref{2.8}, $\rho_n$ is easily calculated in terms of Gamma function
\begin{eqnarray}\label{2.9}
   \rho_n=\bigg(2^n\frac{\Gamma(n+1+\Omega-m)}{\Gamma(1+\Omega-m)}\bigg)^2,
\end{eqnarray}
where $\Omega=(a+b+1)/2$ and $R$ is infinite. The normalization constant $\mathcal{N}(J)$ is given by
\begin{eqnarray}\label{2.10}
   \mathcal{N}^2(J)\equiv\sum_{n=0}^{+\infty}\frac{J^n}{\rho_n}={_1}F_2\Big(1;1+\Omega-m,1+\Omega-m;\frac{J}{4}\Big).
\end{eqnarray}
\indent Using the relation \cite{41}
\begin{eqnarray}\label{2.11}
   \int_0^{+\infty}2J^{\alpha+\beta}K_{\alpha-\beta}(2\sqrt{J})J^{n-1}dJ=\Gamma(n+2\alpha)\Gamma(n+2\beta),\qquad \Re\,n>(-2\,\Re\,\alpha, -2\,\Re\,\beta)
\end{eqnarray}
we find a solution for the moments of a probability distribution \eqref{1.2}
\begin{eqnarray}\label{2.12}
   \rho(J)=\frac{K_0(\sqrt{J})}{2\Gamma^2(1+\Omega-m)}\Big(\frac{J}{4}\Big)^{\Omega-m},
\end{eqnarray}
where $K_0$ being the modified Bessel function. So, the Gazeau-Klauder coherent states \eqref{2.8} finally become
\begin{eqnarray}\label{2.13}
  \Xi^{(m)}(x;J,\gamma)=\frac{\Gamma(1+\Omega-m)}{\sqrt{{_1}F_2\Big(1;1+\Omega-m,1+\Omega-m;\frac{J}{4}\Big)}}
  \sum_{n=0}^{+\infty}\frac{\big(\frac{J}{4}\big)^{n/2}e^{-4i\omega(n-m+\Omega)^2\,t}}{\Gamma(n+1+\Omega-m)}\,\psi_n^{(m)}(x).
\end{eqnarray}
\indent It is easy to see that the Gazeau-Klauder coherent state \eqref{2.13} satisfies the following criteria: (i) continuity of labeling, (ii) temporal stability, (iii) resolution of identity, and (iv) action identity. For example the third condition, associated to the measure $d\mathcal{M}(J,\gamma)$, is related to
\begin{eqnarray}\label{2.14}
  \int|J,\gamma;m\rangle\langle J,\gamma;m|\,d\mathcal{M}(J,\gamma)=\frac{1}{2\pi}\int_{-\pi}^{+\pi}d\gamma\int_0^{+\infty}k(J)|J,\gamma;m\rangle\langle J,\gamma;m|\,dJ,
\end{eqnarray}
if and only if $k(J)$ is defined by
\begin{eqnarray}\label{2.15}
  k(J)= \mathcal{N}^2(J)\rho(J)=\frac{K_0(\sqrt{J})}{2\Gamma^2(1+\Omega-m)}\Big(\frac{J}{4}\Big)^{\Omega-m}{_1}F_2\Big(1;1+\Omega-m,1+\Omega-m;\frac{J}{4}\Big),
\end{eqnarray}
so the resolution of identity is satisfied.

\section{Full and fractional revivals for extended Scarf I potential}%

\noindent In this section, we shall review some of the revival dynamics of the coherent states \eqref{2.13}. It is well understood that systems whose energy spectrum is purely quadratic in the quantum number generate a perfect revivals \cite{33}. In our case, although the energy spectra \eqref{2.7} are not purely quadratic in $n$, it is a simple matter to redefine the quantum number by shifting $n\rightarrow n+m-\Omega$. Then one may suspect the existence of perfect quantum revivals for our system.\\
\indent For general wave-packet, written in atomic units $\hbar=2M=1$,
\begin{eqnarray}\label{3.1}
  |\psi(x,t)\rangle=\sum_{n=0}^{+\infty}c_n e^{-iE_n t}|\varphi(x)\rangle,
\end{eqnarray}
the concept of quantum revivals arises from the weighting probabilities $|c_n|^2$, with $\sum_{n=0}^{+\infty}|c_n|^2=1$. So, when the general wave-packets \eqref{3.1} play the role of our coherent states \eqref{2.13}, then the weighting distribution depends on $J$ as
\begin{eqnarray}\label{3.2}
  |c_n(J)|^2\equiv\frac{J^n}{\mathcal{N}^2(J)\rho_n}=\frac{\Gamma^2(1+\Omega-m)}{{_1}F_2\Big(1;1+\Omega-m,1+\Omega-m;\frac{J}{4}\Big)}
  \frac{\big(\frac{J}{4}\big)^n}{\Gamma^2(n+1+\Omega-m)}.
\end{eqnarray}
\indent In figure 1 we plot the curves $|c_n(J)|^2$ versus $n$ for different values of $J$. We can see that the quasi-Poissonian behavior of weighting distribution is restored at hight values of $J$ and quasi-Gaussian shape are almost localized at right gradually as $m$ becomes more and more bigger.
\begin{figure}[h]
 \centering
 \includegraphics[width=8cm,height=6cm]{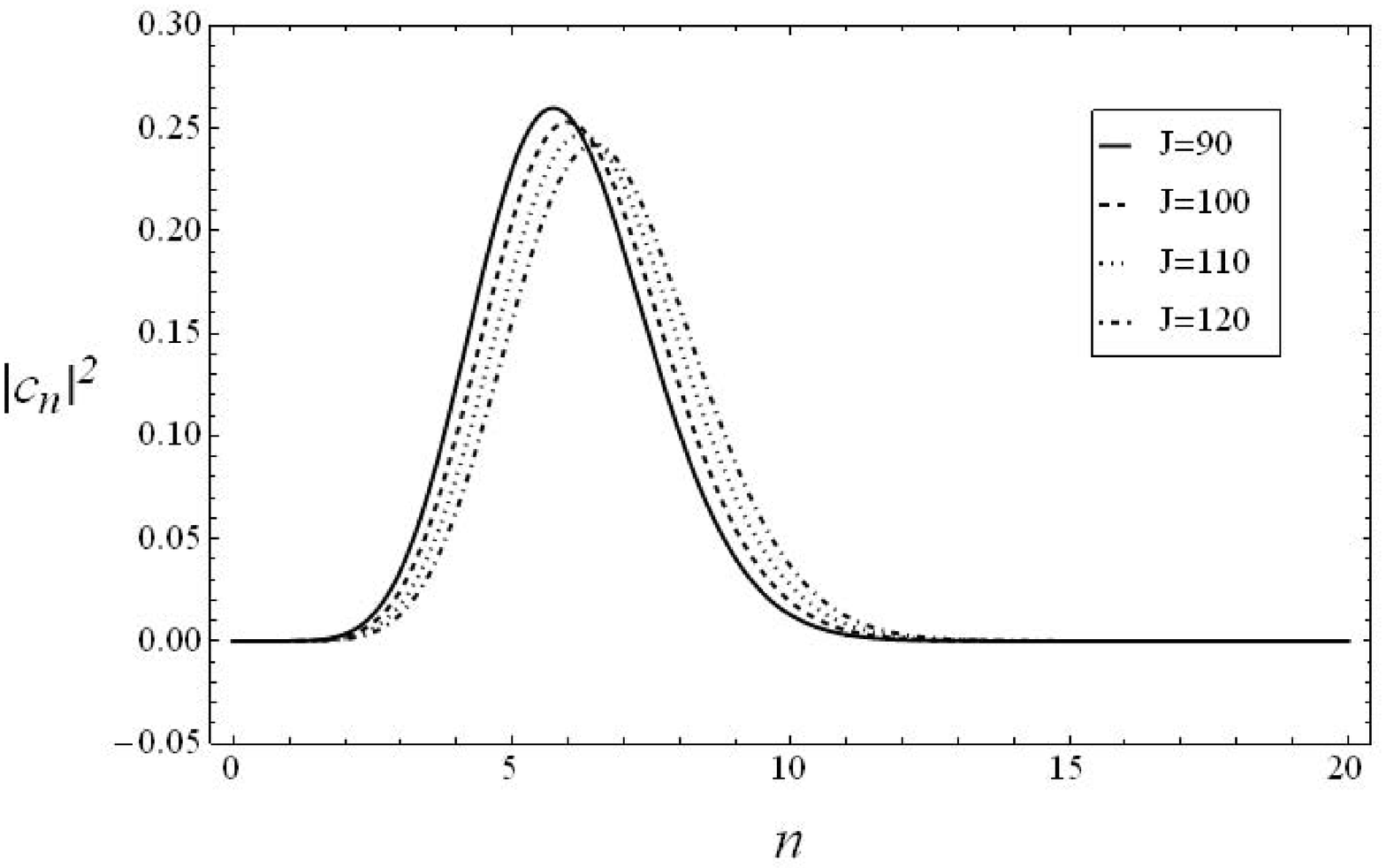}\includegraphics[width=8cm,height=6cm]{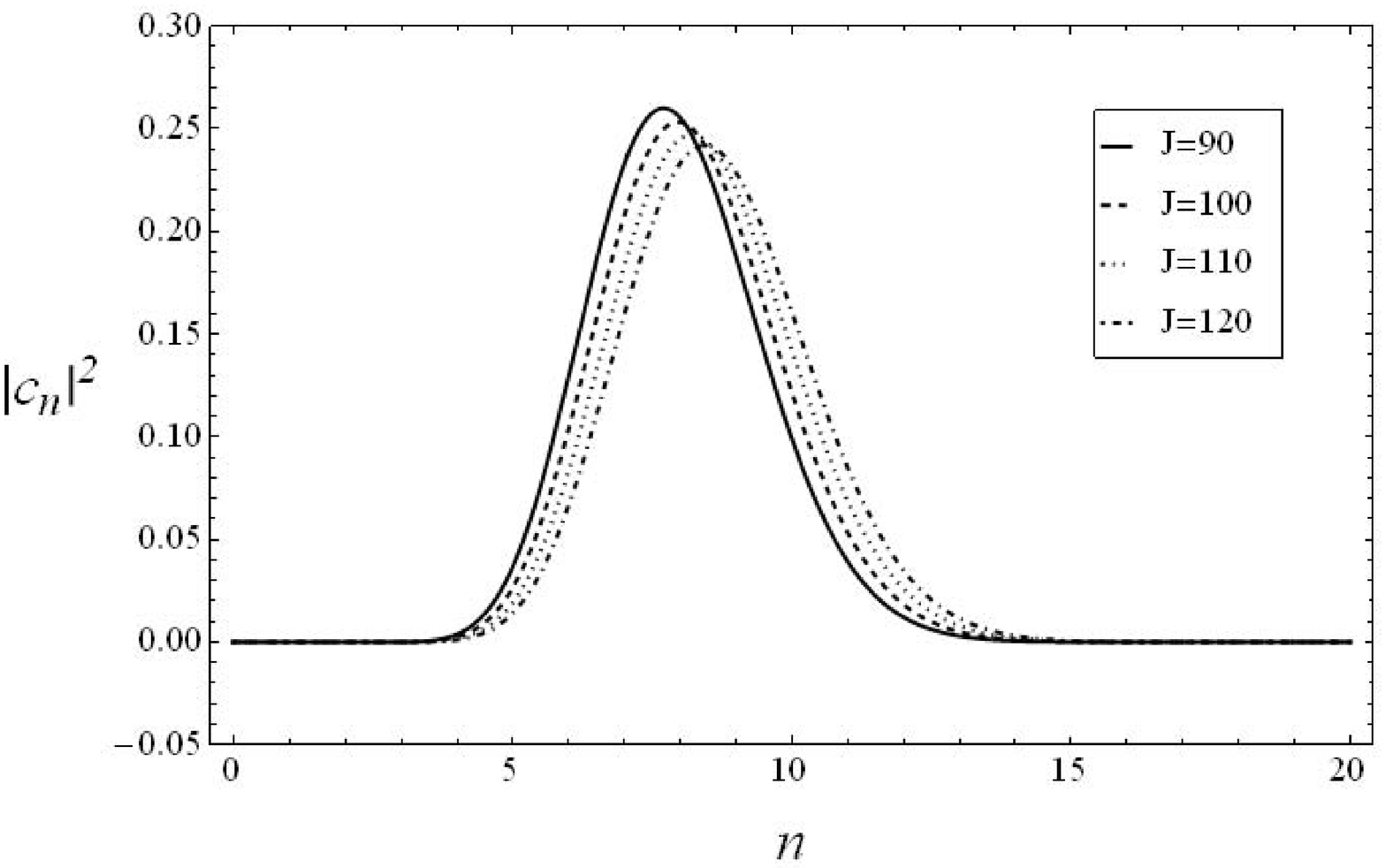}\\
 \caption{The weighting distribution given by \eqref{3.2} for different values of $J$: (left) $a=2.5,\,b=-1/2,\,m=3$ and (right) $a=4.4,\,b=-1/3,\,m=6$.}\label{f1}
\end{figure}
In this case, the mean values of the number operator $\hat N$ are computed
\begin{eqnarray}
  \langle n\rangle&=& \langle J,\gamma;m|\hat N|J,\gamma;m\rangle=\sum_{n=0}^{+\infty}n |c_n(J)|^2
  =\frac{J}{4}\frac{{_1}F_2\Big(2;2+\Omega-m,2+\Omega-m;\frac{J}{4}\Big)}{{_1}F_2\Big(1;1+\Omega-m,1+\Omega-m;\frac{J}{4}\Big)},\label{3.3} \\
  \langle n^2\rangle&=& \langle J,\gamma;m|\hat N^2|J,\gamma;m\rangle=\sum_{n=0}^{+\infty}n^2 |c_n(J)|^2
  =\frac{J}{4}\frac{{_2}F_3\Big(2,2;1,2+\Omega-m,2+\Omega-m;\frac{J}{4}\Big)}{{_1}F_2\Big(1;1+\Omega-m,1+\Omega-m;\frac{J}{4}\Big)},\label{3.4}
\end{eqnarray}
in order to display the Mandel parameter $Q_{\rm M}$ defined by
\begin{eqnarray}\label{3.5}
  Q_{\textrm{M}}=\frac{\langle n^2\rangle-\langle n\rangle^2}{\langle n\rangle}-1.
\end{eqnarray}
\indent The behavior of CS may be characterized through the Mandel parameter \cite{42}. It is an efficient way to characterize non-classical states which have no classical analog. The case of $Q_{\rm M}=0$ coincides with the definition of coherent states, while the cases of $Q_{\rm M}<0$ and $Q_{\rm M}>0$ correspond to the sub-Poissonian (antibunching effect) and super-Poissonian (bunching effect) statistics, respectively. In figure 2 we display the behavior of the Mandel parameter versus $J$ to different states characterizes by the parameters $a,\,b$, and $m$. We note that for a given value of $a$ the state starts with super-Poissonian behavior, and becomes sub-Poissonian from a certain value of $J$ and thereby maintains this state for the rest of the parameter $J$ which hints to a possible squeezing state. We observe, also, that the other states start with super-Poissonian behavior (solid curves) but decrease very fast to sub-Poissonian (dot-dashed curves) as $a$ increases, then we expect that $Q_{\rm M}$ behaves entirely as a sub-Poissonian state from a certain value of $a$, too. Thus both parameters $a$ and $J$ control entirely the evolution of the state.\\
\begin{figure}[]
 \centering
 \includegraphics[width=5.5cm,height=5.3cm]{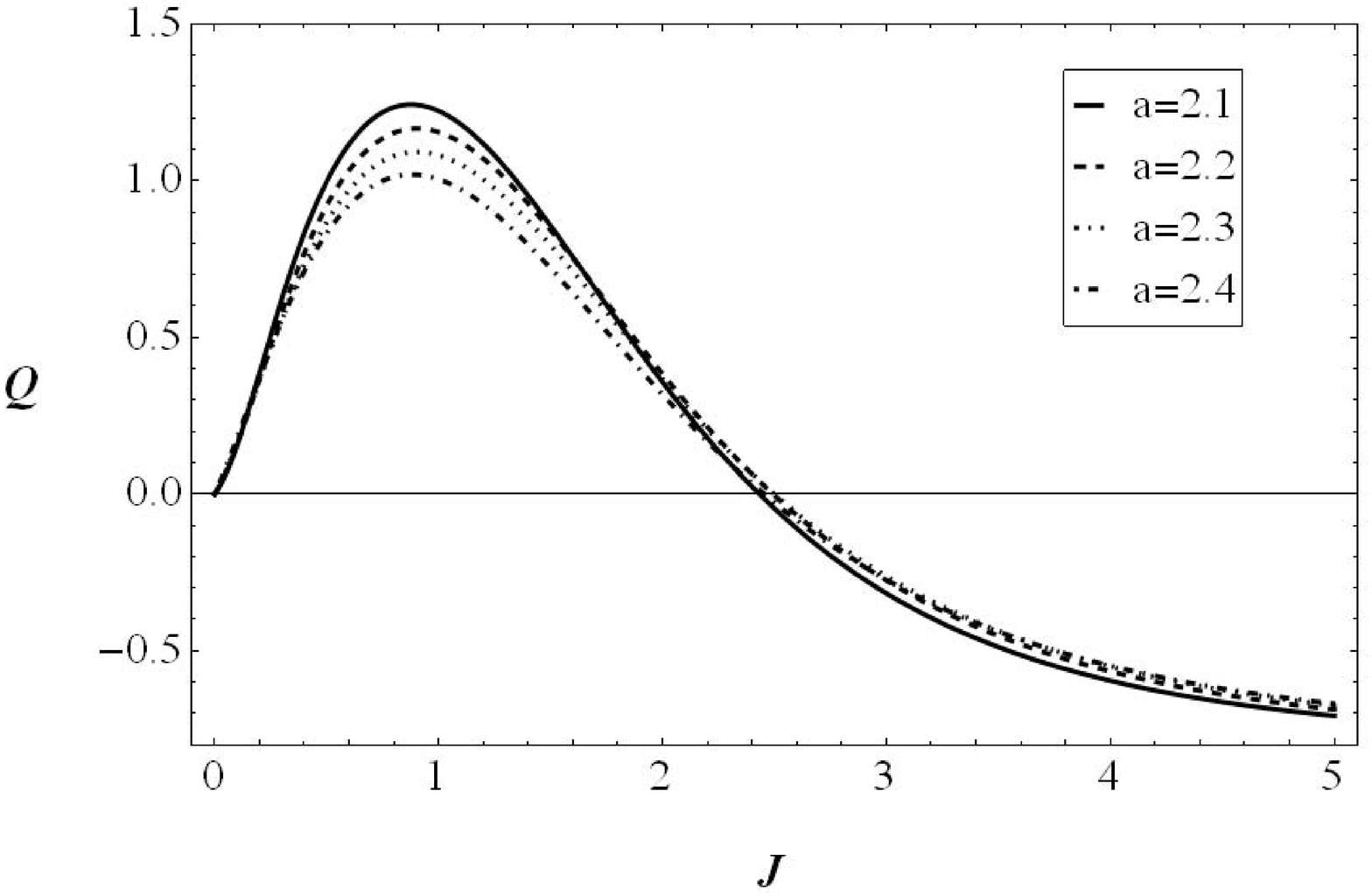}\,\includegraphics[width=5.5cm,height=5.2cm]{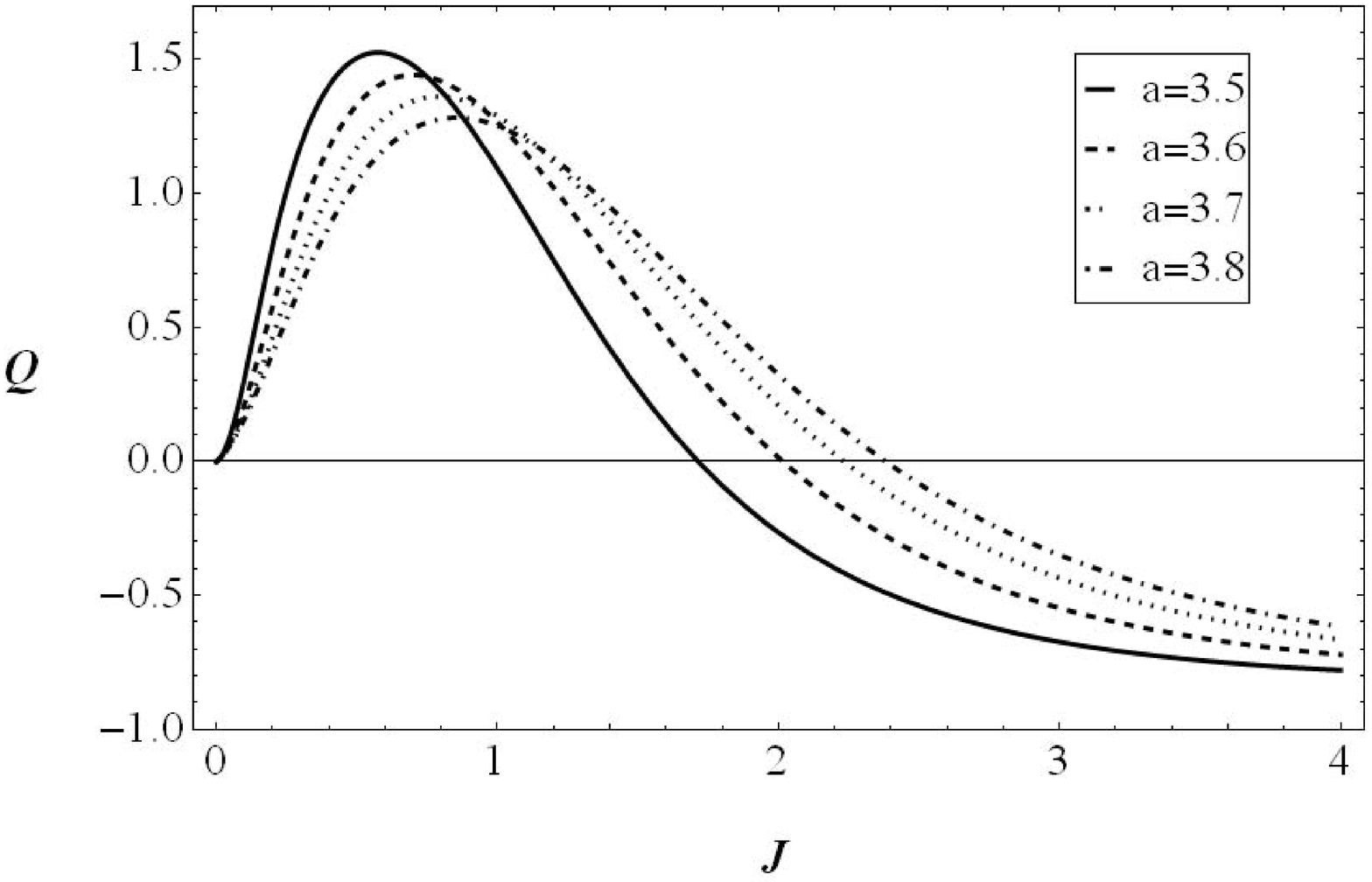}\,\includegraphics[width=5.5cm,height=5.2cm]{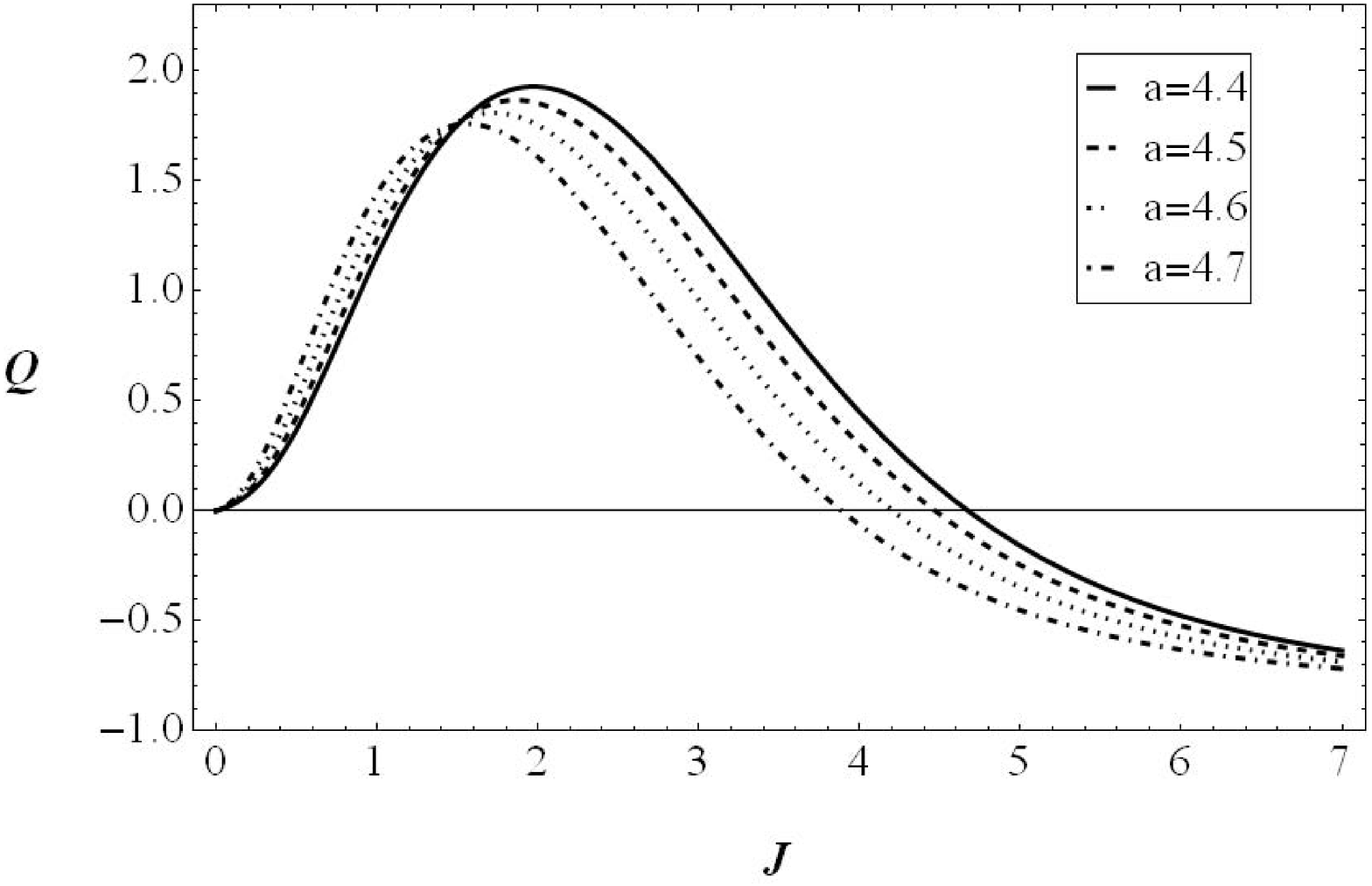}\\
 \caption{The Mandel parameter $Q_{\rm M}$ versus $J$ given in \eqref{3.5} for different values of parameter $a$ and fixed values for $b$ and $m$: (left) $b=-1/2,\,m=4$, (center) $b=-1/4,\,m=5$ and (right) $b=-1/3,\,m=6$.}\label{f2}
\end{figure}
\indent As a prerequisite for obtaining quantum and fractional revivals, we suppose that the expansion \eqref{3.1}, when the wave-packets are precisely our CS, is strongly peaked around a mean value $\langle n\rangle$ such as the spread $\Delta n$ is smaller than $\langle n\rangle\simeq\overline{n}$. This allows us to expand the energy eigenvalues \eqref{2.7} in terms of Taylor series in $n$ around the value $\overline n$ as follows \cite{12,13,14}
\begin{eqnarray}\label{3.6}
  E_n^{(m)}\simeq E_{\overline n}^{(m)}+\frac{dE_n^{(m)}}{dn}\bigg|_{n=\overline n}(n-\overline n)
                  +\frac{1}{2}\frac{d^2E_n^{(m)}}{dn^2}\bigg|_{n=\overline n}(n-\overline n)^2
                  +\frac{1}{6}\frac{d^3E_n^{(m)}}{dn^3}\bigg|_{n=\overline n}(n-\overline n)^3+\cdots,
\end{eqnarray}
where the derivatives of the energy define distinct time scales, namely, the classical period $\Tcl=2\pi/|E_{\overline n}^{(m)}|'$, the revival time $\Trev=2\pi/|\frac{1}{2}E_{\overline n}^{(m)}|''$, the superrevival time $T_{\textrm{sr}}=2\pi/|\frac{1}{6}E_{\overline n}^{(m)}|'''$, and so on. Neglecting the superrevival time, since the energy \eqref{2.7} is a quadratic function in quantum number $n$ and considering up to the second-order term, we can write $E_n^{(m)}$ as
\begin{eqnarray}\label{3.7}
  E_n^{(m)}=4\,\omega\Big[(\overline n-m+\Omega)^2+2(\overline n-m+\Omega)(n-\overline n)+(n-\overline n)^2\Big],
\end{eqnarray}
which the timescales are define by
\begin{eqnarray}\label{3.8}
  \Tcl=\frac{\pi}{4\,\omega\,(\overline n-m+\Omega)}\qquad\textrm{and}\qquad\Trev=\frac{\pi}{2\,\omega}\qquad\Rightarrow\qquad\Trev=2(\overline n-m+\Omega)\Tcl.
\end{eqnarray}
\indent In the present situation, using the last definitions, the Gazeau-Klauder CS reads (in units of $\overline t=t/\Tcl$) as
\begin{eqnarray}\label{3.9}
  |\Xi^{(m)}(x,\overline t)\rangle=e^{-i\pi(\overline n-m+\Omega)\overline t}\sum_{n=0}^{+\infty}c_n(J)
  \exp\Big\{\frac{-i\pi(n-\overline n)(n+\overline n-2m+2\Omega)}{\overline n-m+\Omega}\overline t\Big\}|\psi_n^{(m)}(x)\rangle,
\end{eqnarray}
where $c_n(J)$ are defined by \eqref{3.2}. For the coherent states \eqref{3.9} it is possible to calculate the quantum revivals through the autocorrelation function defined as the projection of a wave-packet CS onto its initial state
\begin{eqnarray}\label{3.10}
  A(\overline t) &=& \langle\Xi^{(m)}(x,0)|\Xi^{(m)}(x,\overline t)\rangle\nonumber \\
       &=& \frac{e^{-i\pi(\overline n-m+\Omega)\overline t}\Gamma^2(1+\Omega-m)}{{_1}F_2\Big(1;1+\Omega-m,1+\Omega-m;\frac{J}{4}\Big)}
       \sum_{n=0}^{+\infty}\frac{\exp\Big\{\frac{-i\pi(n-\overline n)(n+\overline n-2m+2\Omega)}{\overline n-m+\Omega}\overline t\Big\}}{\Gamma^2(n-m+\Omega+1)}\bigg(\frac{J}{4}\bigg)^n.
\end{eqnarray}
\indent Using the relation \cite{40}
\begin{eqnarray}\label{3.11}
  \sum_{n=0}^{+\infty}c_nx^n=\bigg(\sum_{n=0}^{+\infty}a_nx^n\bigg)\bigg(\sum_{n=0}^{+\infty} b_nx^n\bigg),
  \qquad\textrm{with}\qquad c_n=\sum_{k=0}^n a_k\,b_{n-k},
\end{eqnarray}
the straightforward calculations lead to express the squared modulus of $A(\overline t)$ as
\begin{eqnarray}\label{3.12}
  |A(\overline t)|^2=\frac{\Gamma^4(1+\Omega-m)}{{_1}F_2^2\Big(1;1+\Omega-m,1+\Omega-m;\frac{J}{4}\Big)}
       \sum_{n=0}^{+\infty}\sum_{k=0}^n\frac{\exp\Big\{\frac{i\pi(n-2k)(n-2m+2\Omega)}{\overline n-m+\Omega}\overline t\Big\}\big(\frac{J}{4}\big)^n}{\Gamma^2(k-m+\Omega+1)\Gamma^2(n-k-m+\Omega+1)},
\end{eqnarray}
where in order to make numerical calculations we need to truncate the series at some order in $n$, keeping in mind that $n\geq m$.\\
\begin{figure}[t]
 \centering
 \includegraphics[width=8.5cm,height=5cm]{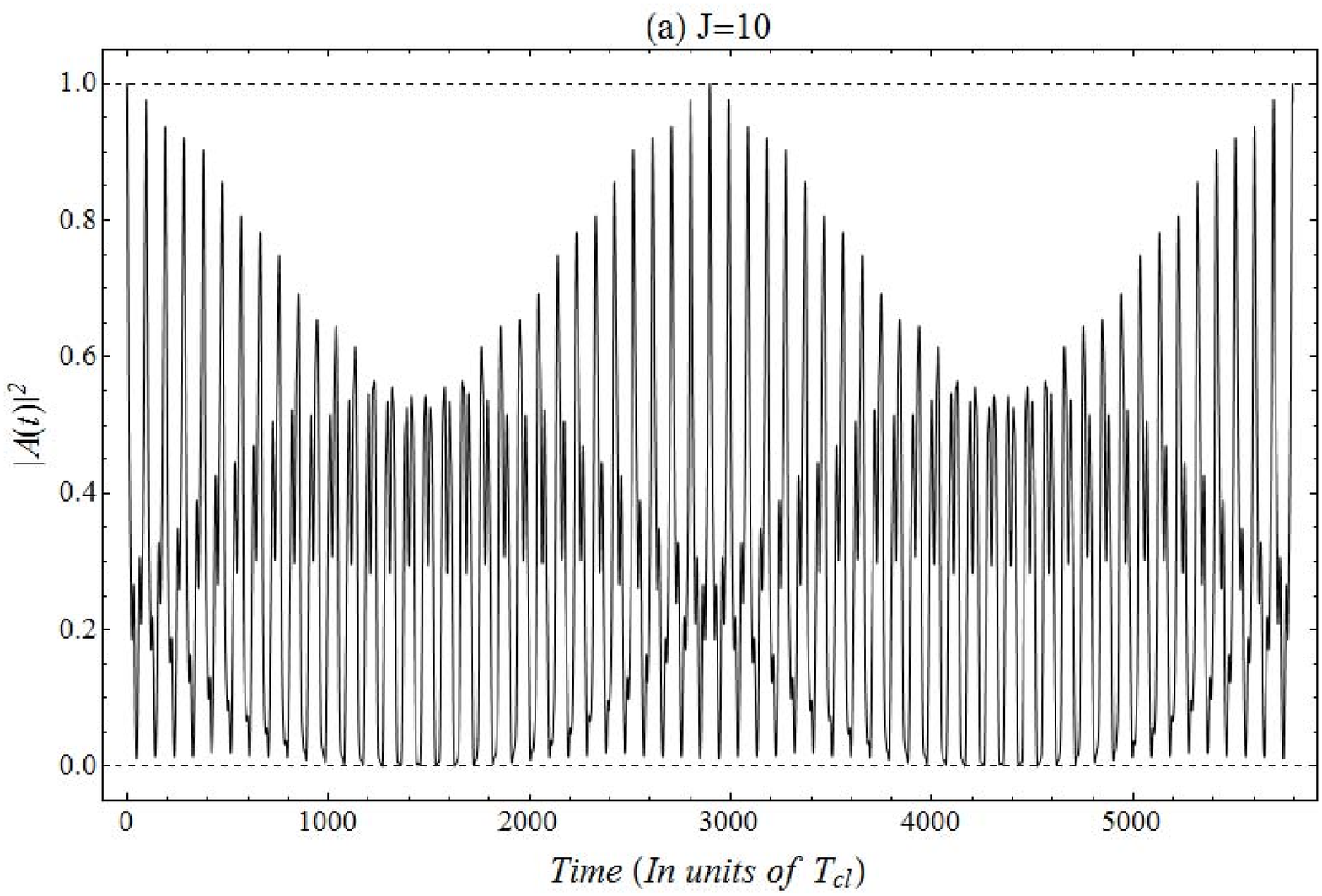}\,\,\,\,\includegraphics[width=8.5cm,height=5cm]{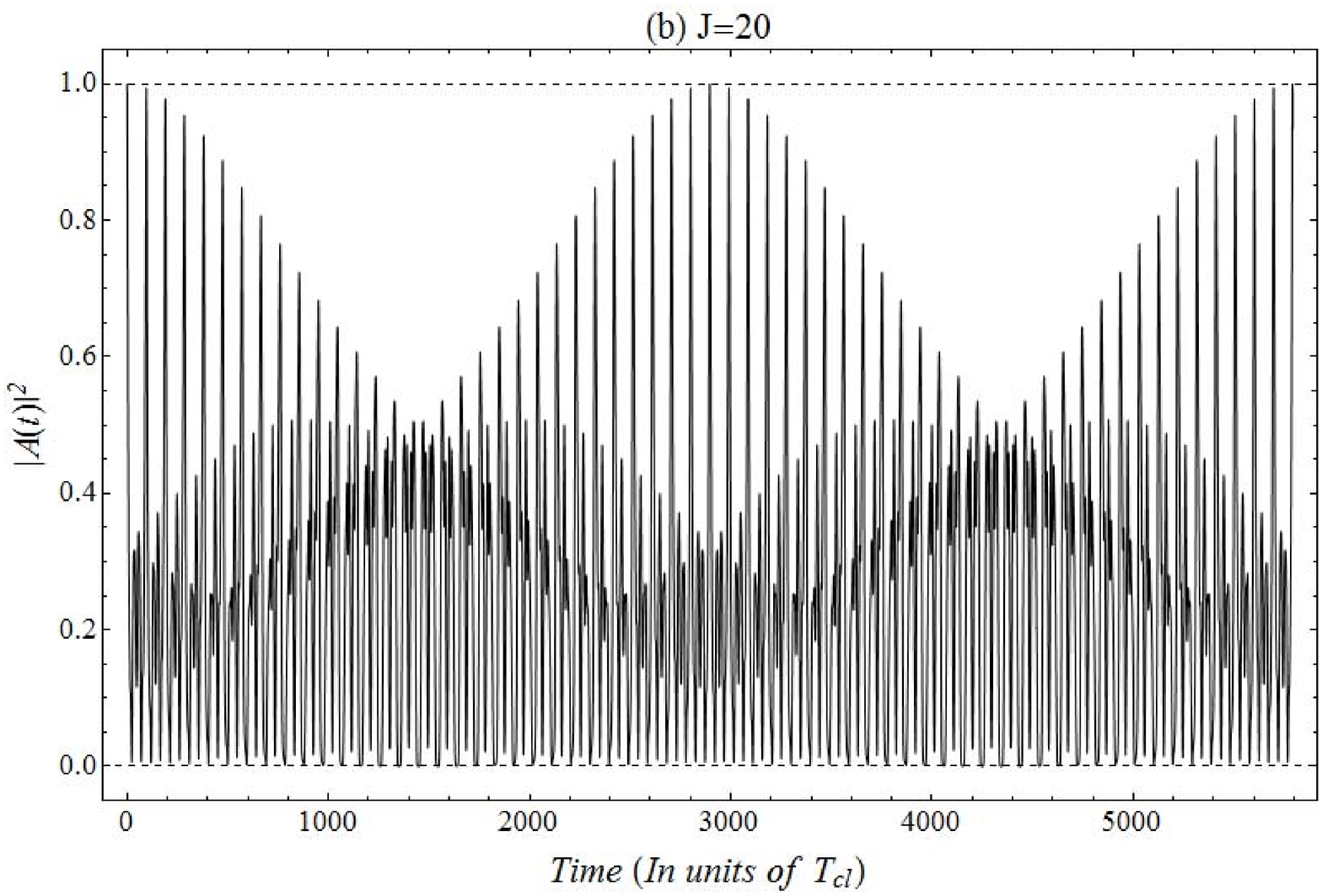}\\[2mm]
 \includegraphics[width=8.5cm,height=5cm]{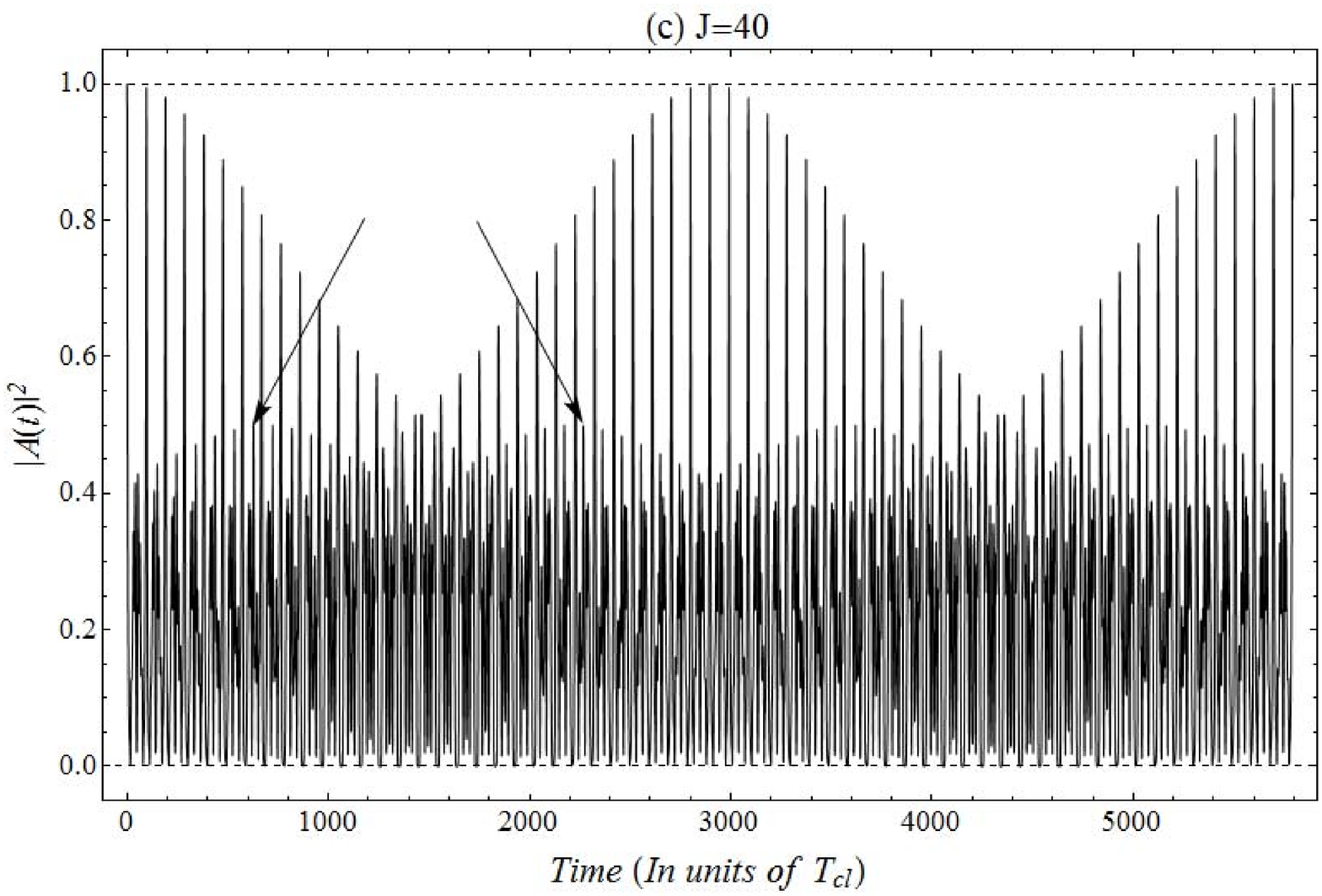}\,\,\,\,\includegraphics[width=8.5cm,height=5cm]{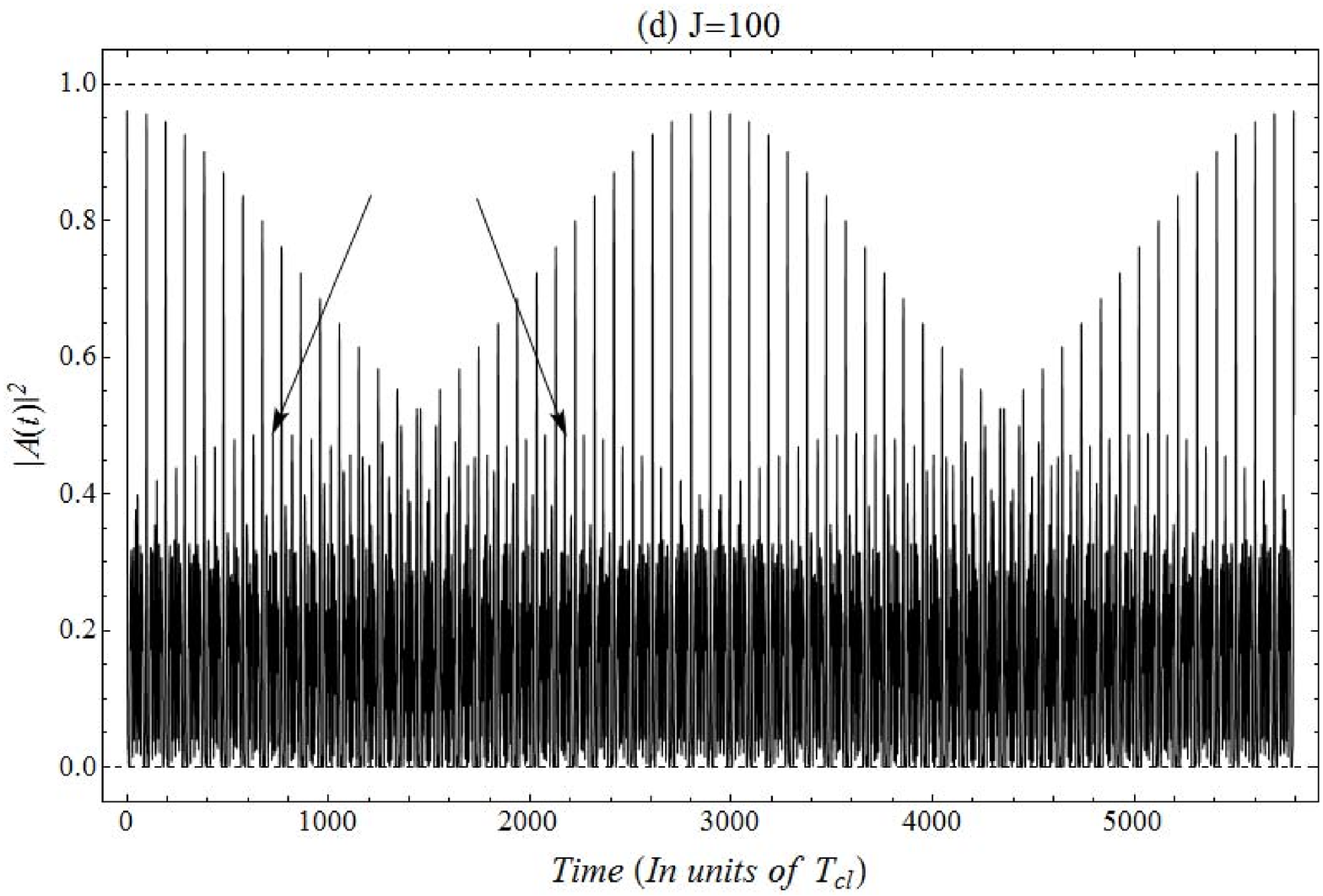}\\
 \caption{The squared modulus of the autocorrelation function plotted from \eqref{3.12} along $t=2\,\Trev$ for $a=4.4,\,b=-1/3$ and $m=6$. The time $t$ is given in units of $\Tcl$.}\label{f3}
\end{figure}
\indent The evolution of $|A(\overline t=t/\Tcl)|^2$ is shown in figure 3 in the cases $J=10,\,20,\,40$ and $100$. These plots are generated by choosing $n_{\textrm{max}}=50$ and $\overline n=100$ for $a=4.4,\,b=-1/3$ and $m=6$. The revival time calculated through \eqref{3.12} is $\Trev\simeq2896.825$ (in units of $\Tcl$) and using \eqref{3.8} one find $\Tcl\simeq15$. All cases clearly show that the time evolution of the system is periodic with period $\Trev$, which is an exact revival time too, since the revival structures are not modulated by a superrevival time. The sharp peaks appearing in the cases $J=20,\,40$ and 100 are characterize by the occurrence of fractional revivals which become more and more apparent as $J$ increases. The function \eqref{3.12} proves to be symmetric about $\Trev/2$ which means that all the features of coherent wave-packet evolution appear at a time interval from 0 to $\Trev/2$. After that time all the events will repeat again and again. As we can see in figures 3(c) and (d) especially interesting are fractional revivals occurring at quarters of the revival time (indicated by an arrow) showing a suspect symmetry which indicates a strong occurrence of fractional revivals at that times since the energy is quadratic in $n$. Unfortunately these peaks are uniformly mixed so that we do not have any idea on which value of the autocorrelation function appears at what time; thus fractional revivals are still invisible and should be resolved. Then a natural question arises: what about the reconstruction and location of fractional revivals? This question will be considered in more details in a forthcoming section.

\section{Reconstruction and location of fractional revivals from wavelet-based time-frequency representation}%

\noindent In this section our key aim is the observation of fractional revivals of the potential \eqref{2.3} since the autocorrelation function fails to reconstruct them. An efficient and especially illustrative method for this kind of construction involves the use of the time-frequency analysis based on the continuous wavelet transform (CWT) \cite{34,35,36}; this latter is basically a two-dimensional function of the scale $s$ (inversely to the frequencies) and the time shift $\tau$. The specific choice of using CWT rather than short-time Fourier transform (STFT) consists in the fact that STFT uses a single analysis window, while CWT uses short windows at high frequencies and long windows at low frequencies. This property overcomes the limitation of STFT and allows us to uncover finer details of signal acting at smaller scales.\\
\indent Here, in our case, the signal is our autocorrelation function given by \eqref{3.12} and the kernel function is the well-known Morlet wavelet defined by
\begin{eqnarray}\label{4.1}
  h(t)=\frac{1}{\pi^{1/4}}\,e^{i\omega_0t}\,e^{-t^2/2},
\end{eqnarray}
where $\omega_0$ is called the central frequency. Let us now be more specific by redefining $t\rightarrow\zeta=(t-\tau)/s$, the CWT of $|A(t)|^2$ given in \eqref{1.5} can be written as
\begin{eqnarray}\label{4.2}
  W_{|A|^2}(s,\tau)=\sqrt{s}\int_{-\infty}^{+\infty}|A(s\zeta+\tau)|^2h^{\ast}(\zeta)\,d\zeta,
\end{eqnarray}
where $h(\zeta)=\pi^{-1/4}e^{i\omega_0\zeta}e^{-\zeta^2/2}$ and by performing the integration over $\zeta$ we get
\begin{eqnarray}\label{4.3}
  W_{|A|^2}(s,\tau)&=&\pi^{-1/4}\sqrt{2\pi s}\sum_{n=0}^{+\infty}\sum_{k=0}^{+\infty}|c_n(J)|^2|c_k(J)|^2
  \exp\bigg\{-2\pi i(n-k)\bigg(\frac{1}{\Tcl}+\frac{n+k-2\,\overline n}{\Trev}\bigg)\tau\bigg\}\nonumber\\
  &&\times\exp\bigg\{-\frac{1}{2}\bigg[\omega_0+2\pi s(n-k)\bigg(\frac{1}{\Tcl}+\frac{n+k-2\,\overline n}{\Trev}\bigg)\bigg]^2\bigg\}.
\end{eqnarray}
\indent Despite we do not know an exact expression for the central frequency $\omega_0$, it is useful to identify, at first approximation, $\omega_0\simeq\omega=k^2/4$ in order to calculate the minima of $W_{|A|^2}(s,\tau)$. For convenience, we reproduce here, in part, some of results obtained by Ghosh and Banerji \cite{15}. Since the weighting functions $|c_n(J)|^2$ and $|c_k(J)|^2$ appearing in \eqref{4.3} are peaked around $\overline n$, the central frequency can be calculated by replacing $n$ by $\overline n$ and $k$ by $\overline n+p$, where $p$ is a positive integer. This gives rise to the relation
\begin{equation}\label{4.4}
    \omega=\frac{2\pi s p}{\Tcl}\bigg(1+p\frac{\Tcl}{\Trev}\bigg)\sim\frac{2\pi s}{\Tcl}\,p,\qquad\qquad\bigg(\frac{\Tcl}{\Trev}\ll1\bigg),
\end{equation}
and since the scale parameter $s$ is related to frequency $\nu$ by the relation $s=\omega/(2\pi\nu)$, then we obtain the formulae
\begin{equation}\label{4.5}
    \nu\simeq\frac{p}{\Tcl},\qquad\textrm{and}\qquad s=\frac{\omega}{2\pi\nu}.
\end{equation}
\indent Each frequency corresponds a particular time $\tau$ given by
\begin{equation}\label{4.6}
    \tau=\frac{q}{2p}\,\Trev,
\end{equation}
where $q$ is a positive integer. If the restriction $q<2p$ holds, then one can see the correspondence between \eqref{4.6} and the well-defined fractional revivals of the autocorrelation function $A(t)$ localized at various rational multiples of the revival time given by $t=(r/l)\,\Trev\,(\textrm{mod}\,\,l)$. Both expressions \eqref{4.5} and \eqref{4.6} provide the required resolution in frequency and the desired location in time for the reconstruction and location of fractional revivals; in other words, one can find both time and frequency in terms of the two timescales $\Trev$ and $\Tcl$.\\
\indent Ghosh and Banerji used their results to correctly analyze the localization of different patch corresponding to the fractional revival it contributes to, using the \textit{Time-Frequency ToolBox for MatLab} \cite{43}. Here we test the robustness and efficiency of CWT as a mathematical filter, in order to filter out fractional revivals from the complicated plots of figure 3 obtained from the autocorrelation function. To this end we restrict our analysis to the reconstruction and localization of fractional revivals. We illustrate this by plotting in figure 4 the squared modulus of both $A(t)$ and $W_{|A|^2}(\tau)$. The figure (4a) displays the evolution of $|A(t)|^2$ given by \eqref{3.12} in units of $\Tcl$ along $t=10\,\Trev$, while figure (4b) gives the time dependence of $|W_{|A|^2}(\tau)|^2$ given by \eqref{4.3} in units of $\Trev$ for the first harmonic ($p=1$) and at a time interval from $2\,\Trev/9$ to $2\,\Trev/7$, i.e., near the first quarter revival $t\simeq\Trev/4$. As shown on the figure (4b), the coherent wave-packets evolve in time and split into a collection of spatially distributed mini-"\textit{empire state building}" groups, each of them closely reproduces the shape of the initial coherent wave-packets of figure (4a), and are all of nearly constant values. The motion of these fractional revivals is periodic with period given by a rational fractions of $\Tcl\simeq15$ calculated from \eqref{3.8}, as can be seen on the figure (4b). Thus, for our system under study, we can conclude that the wave packet remains coherent for a number of classical periods.\\
\begin{figure}[h]
 \centering
 \includegraphics[width=16cm,height=7.5cm]{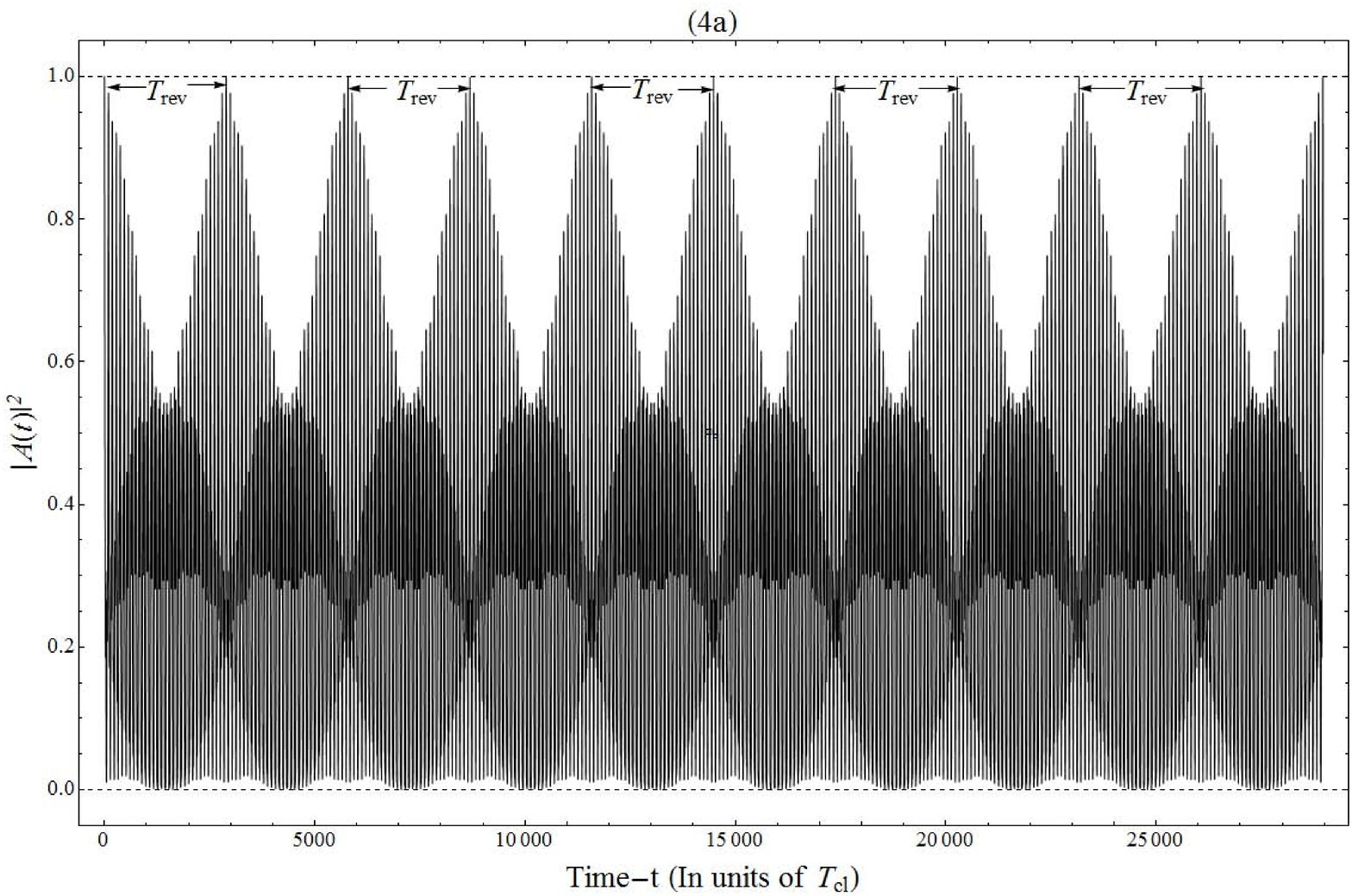}\\[3mm]\includegraphics[width=16cm,height=7.5cm]{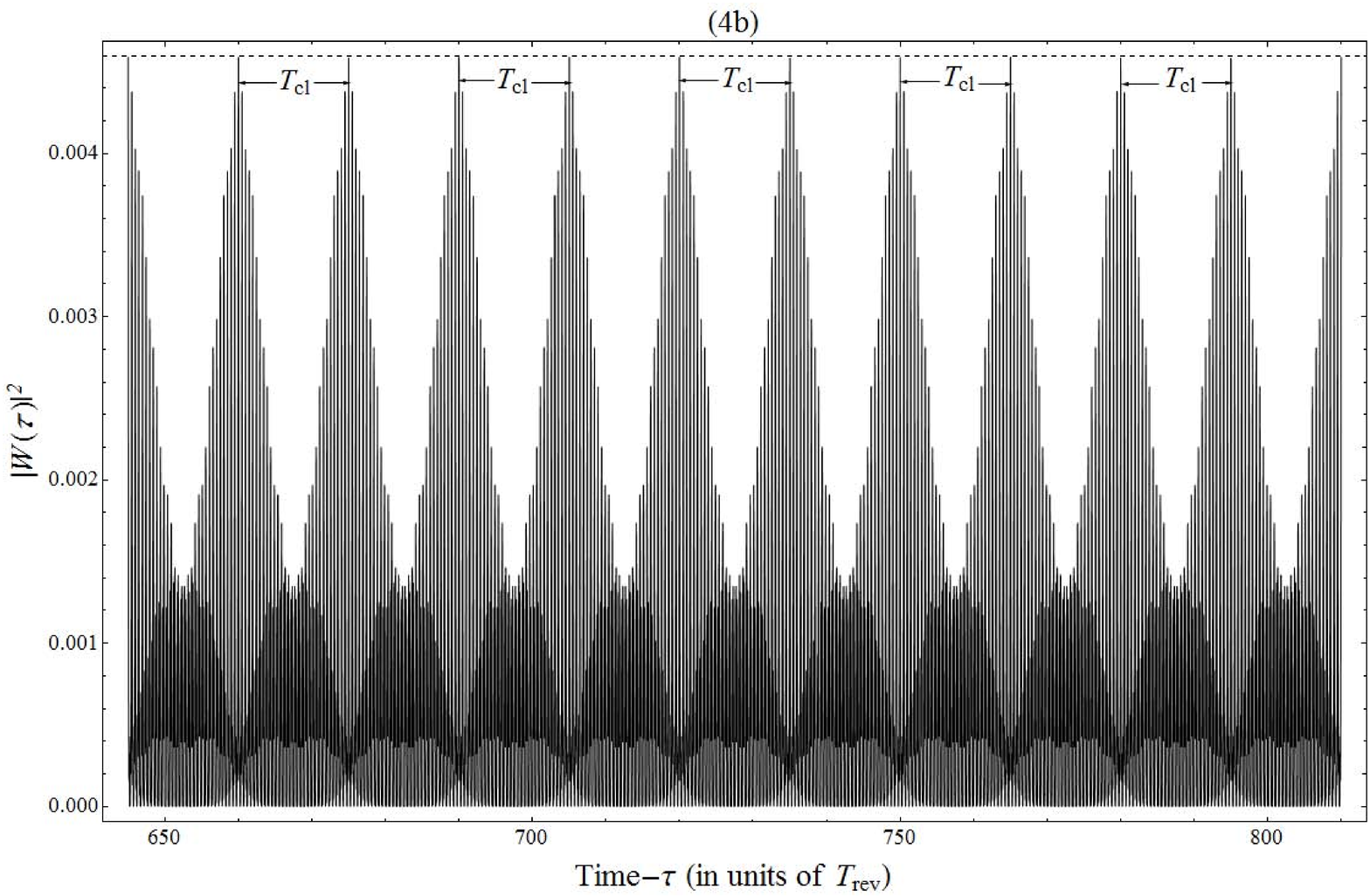}\\
 \caption{(4a) The squared modulus of the autocorrelation function $A(t)$ defined by \eqref{3.12} along $t=10\,\Trev$, (4b) the squared modulus of $W_{|A|^2}(\tau)$ defined by \eqref{4.3} for $p=1$. Both figures are plotted for parameters: $a=4.4,\,b=-1/3,\,m=6$ and $J=10$.}\label{f4}
\end{figure}
\indent Now to correctly analyze the location of fractional revivals, we use \eqref{4.5} and \eqref{4.6} to express $W_{|A|^2}(s,\tau)$ in terms of the parameters $p$ and $q$, which gives
\begin{eqnarray}\label{4.7}
  W_{|A|^2}(p,q)&=&\pi^{-1/4}\sqrt{\frac{\omega\Tcl}{p}}\sum_{n=0}^{+\infty}\sum_{k=0}^{+\infty}|c_n(J)|^2|c_k(J)|^2
  \exp\bigg\{-2\pi i(n-k)\bigg(\frac{\Trev}{\Tcl}+n+k-2\,\overline n\bigg)\frac{q}{2p}\bigg\}\nonumber\\
  &&\times\exp\bigg\{-\frac{\omega^2}{2}\bigg[1+\frac{n-k}{p}\bigg(1+(n+k-2\,\overline n)\frac{\Tcl}{\Trev}\bigg)\bigg]^2\bigg\},
\end{eqnarray}
\begin{figure}[h]
 \centering
 \includegraphics[width=10cm,height=7.2cm]{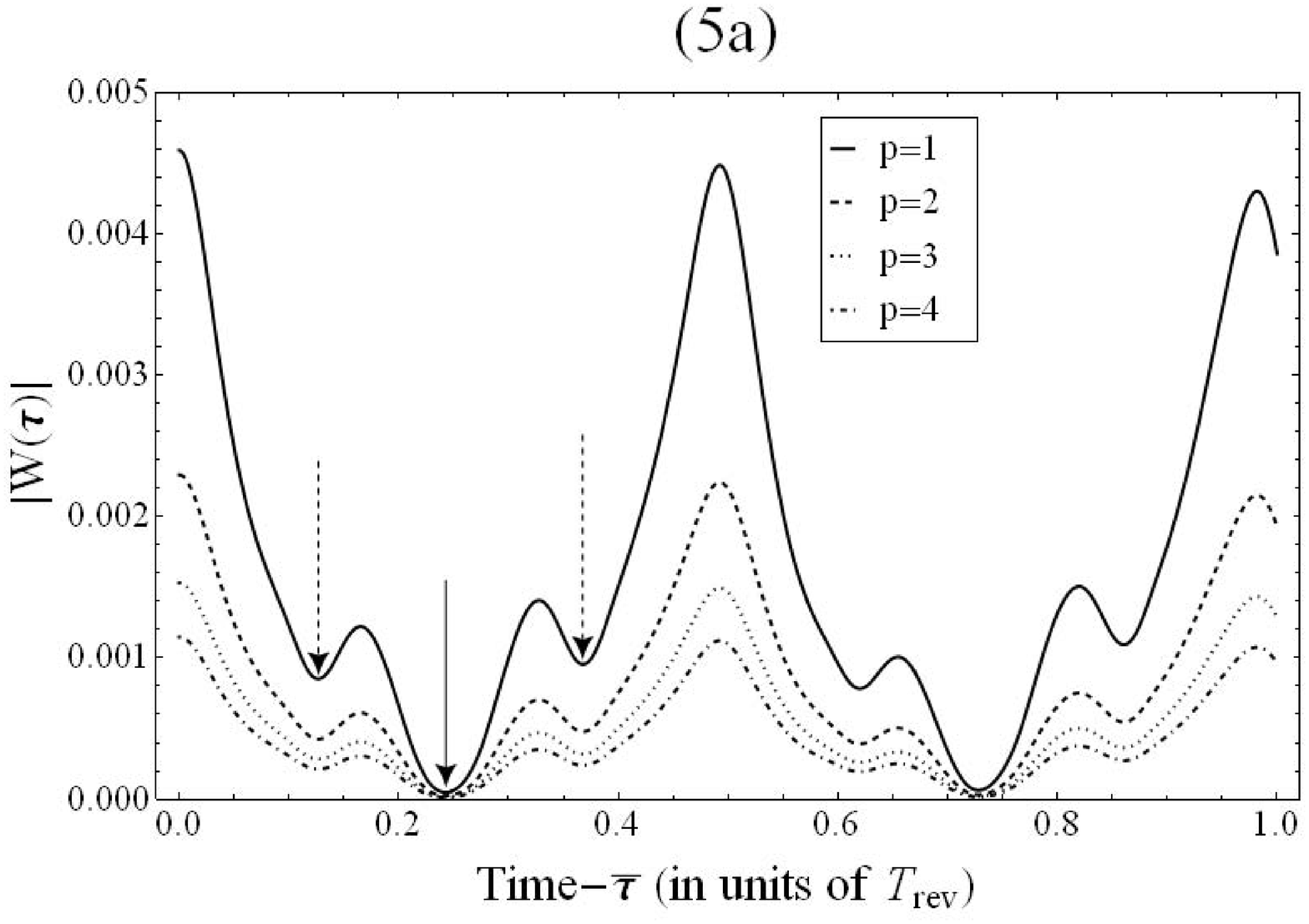}\includegraphics[width=7cm,height=7cm]{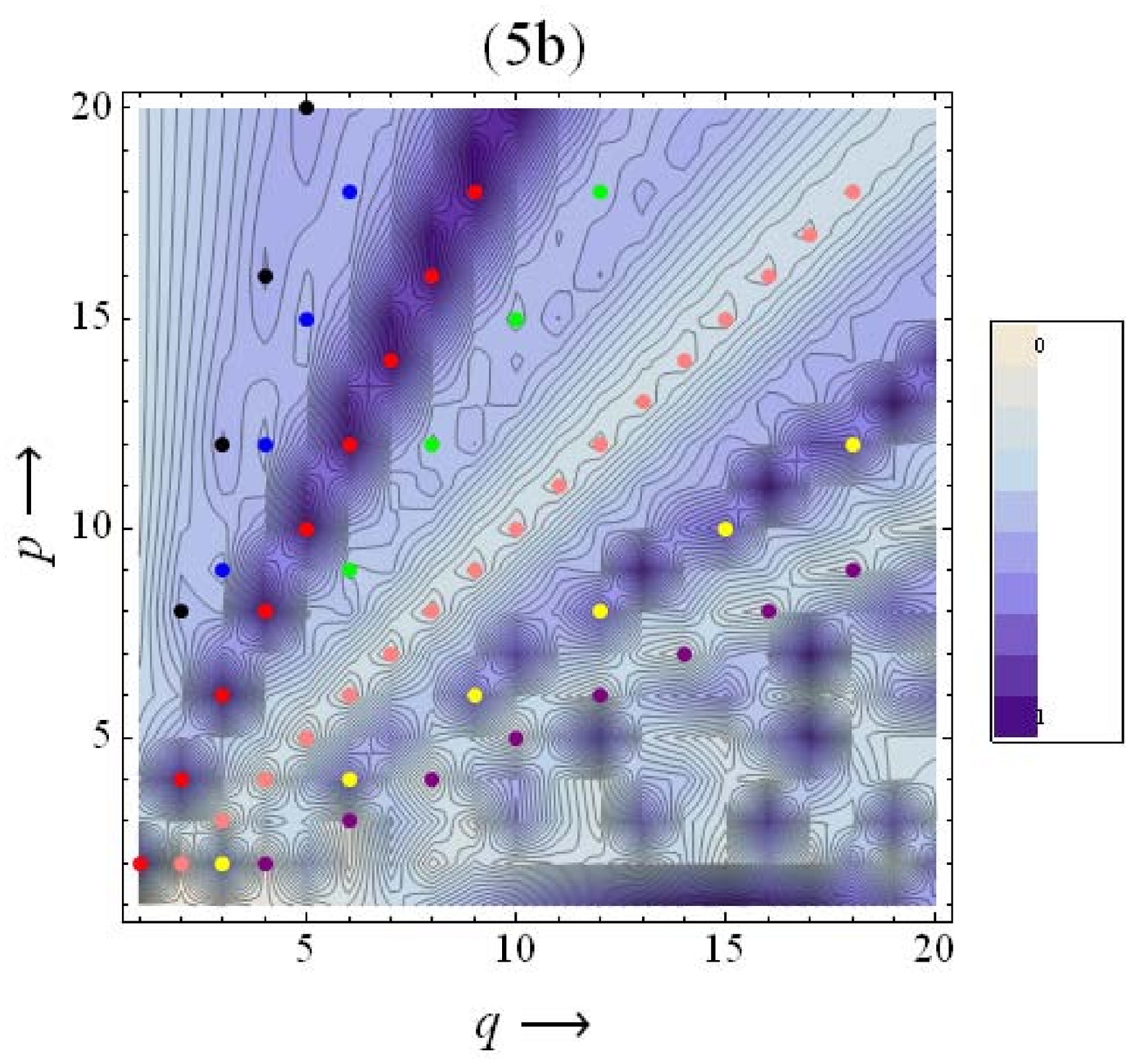}\\
 \caption{(5a) The plot of the CWT of the autocorrelation function defined by \eqref{4.3} in units of $\overline\tau=\tau/\Trev$, (5b) log density plot of the squared modulus of \eqref{4.7}.}\label{f5}
\end{figure}
and we display in figure 5 the CWT of the autocorrelation function \eqref{4.3} and the density plot of $\ln(|W_{|A|^2}(p,q)|^2)$, for different values of $p$ and $q$. The figure (5a) shows how CWT of the autocorrelation function $A(t)$ displays a possible location of fractional revivals at both absolute minimum (vertical arrow) and relative minima (dashed arrows) of $W_{|A|^2}(\tau)$, for the first four harmonics $p=1,2,3$, and $4$. The absolute minimum clearly predicts that the first quarterly fractional revival should occur around $\overline\tau=\tau/\Trev\simeq1/4$, and fade away for $\overline\tau>1/4$. One can observe that they are more better localized than those occurring at vicinity of $\overline\tau\simeq3/4$. These imperfect and asymmetric quarter revivals take the slightly modified form due essentially to the fact that the energy spectrum are not purely quadratic in the quantum number $n$. However, the relatively minimum areas can be the scene of several other orders of fractional revivals but with a less significant probability amplitude than those detected in the absolute minimum areas.\\
\indent Indeed, referring to the figure (5b), one can observe that the occurrence of fractional revivals is clearly better indicated by $\ln(|W_{|A|^2}(p,q)|^2)$ than by $|A(t)|^2$. A characteristic straight dark rays are clearly visible on the $p-q$ plan, which correspond to the different areas (patches) where the fractional revivals occur and denoted by the different colored points. Each point is well-localized about a particular $p$ (i.e., frequencies $\nu$) and $q$ (i.e., time $\tau$) and located at the center of the corresponding patch; it tells us which the fractional revival it contributes to. For examples, using \eqref{4.6}, we can observe that the red points are all localized at the first quarter revival, i.e., $\tau=\Trev/4$, while the corresponding value of time associated to the yellow points is localized at the second one, i.e., at $\tau=3\,\Trev/4$. Thus the evolution of the system shows clearly the occurrence of fractional revivals exactly at quarters of the revival time, with a little imperfect and asymmetric modified form of rays due to the raison mentioned above. However the pink and purple points, both localized on the white regions, are all centered at times $\tau=\Trev/2$ and $\tau=\Trev$, respectively. This indicates clearly that one does not observe the occurrence of the fractional revivals at these times.\\
\indent On the other hand, the existence of relatively minimum areas clearly manifest a slight emergence of a few patches showing the location of fractional revivals at the time $\tau=\Trev/8$ (black points), $\tau=\Trev/6$ (blue points), and $\tau=\Trev/3$ (green points), respectively, and disappeared completely in the interval $\Trev/2$ and $\Trev$, except of course for those associated to $\tau=3\,\Trev/2$ (yellow points). These occurrences are restrictive in the sense that are localized in relatively minimum areas, as displays in figure 5(a), with a certain probability densities less than those attributed to patches located at exactly quarters of the revival time.

\section{Conclusion}%

\noindent In this paper, we have constructed the Gazeau-Klauder CS for the extended Scarf I potential associated with exceptional $X_m$ Jacobi OP. These CS satisfy the criteria of continuity of labeling, resolution of unity, temporal stability, and action identity. The probability distribution shows a Gaussian behavior for a large values of $J$. The Mandel parameter starts with super-Poissonian behavior and decreases very fast to sub-Poissonian as $a$ and $J$ increase and remains in this state for the rest of these parameters, which indicates that our CS \eqref{3.9} exhibit squeezing.\\
\indent Using the autocorrelation function of \eqref{3.9}, a full revivals are obviously occur at each multiple of $\Trev$ and sharp peaks revel a little signature of the fractional revival structure. These latter are uniformly mixed, so that their orders can not be determined exactly. To overcome this problem, we have made used of the continuous wavelet transform of $A(t)$ in order to reconstruct and localize fractional revivals. This study has shown (i) that fractional revivals faithfully reproduce the shape of the initial coherent wave-packets, within an imperfect and asymmetric quarterly form due to the fact that the energy spectrum are not purely quadratic in $n$, and (ii) the location of fractional revivals is clearly better viewed by $\ln(|W_{|A|^2}(p,q)|^2)$ than by $|A(t)|^2$, which indicates that the wavelet-based time-frequency representation can be regarded as a powerful tool and a complementary method for reconstructing and localizing fractional revivals.


\end{document}